\documentstyle[12pt]{article}
\newcommand{\mc}{\multicolumn}
\newcommand{\bce}{\begin{center}}
\newcommand{\ece}{\end{center}}
\newcommand{\beq}{\begin{equation}}
\newcommand{\eeq}{\end{equation}}
\newcommand{\bea}{\vspace{0.25cm}\begin{eqnarray}}
\newcommand{\eea}{\end{eqnarray}}

\newcommand{\bsigma}{\mbox{\boldmath $\sigma$}}
\newcommand{\btau}{\mbox{\boldmath $\tau$}}

\newcommand{\bj}{{\bf j}}

\newcommand{\br}{{\bf r}}

\newcommand{\ba}{\begin{array}}
\newcommand{\ea}{\end{array}}

\newcommand{\ie}{{\sl i.e.~}}

\newcommand{\ket}[1]{| {#1} \rangle}
\newcommand{\bra}[1]{\langle {#1} |}

\def\lsim{\mathrel{\rlap{\lower4pt\hbox{\hskip1pt$\sim$}}
    \raise1pt\hbox{$<$}}}	  
\def\gsim{\mathrel{\rlap{\lower4pt\hbox{\hskip1pt$\sim$}}
    \raise1pt\hbox{$>$}}}	  

\begin{document}
\vspace{1.0in}
\begin{flushright}
{\small February 25, 1993}
\end{flushright}
\vspace{2.0cm}
\bce
{\large{\bf M1 RESONANCES IN UNSTABLE MAGIC NUCLEI}}
\vspace{.55in}

S. KAMERDZHIEV $\,$\footnote{Permanent address: Institute of
Nuclear and Power Engineering, 249020 Obninsk, Russia},
 J. SPETH $\,$\footnote{also at: Institut f\"ur theoretische Kernphysik,
 W-5300 Bonn, FRG}, G. TERTYCHNY$^{\scriptsize 1}$\\
and J. WAMBACH $\,$\footnote{also at: Dept. of Physics,
University of Illinois at Urbana-Champaign, Urbana, IL 61801, USA}

\vspace{1.0cm}
{\it Institut f\"ur Kernphysik\\
Forschungszentrum J\"ulich\\
W-5170 J\"ulich, FRG}
\vspace{1.0cm}
\ece

\vspace{.65in}
\begin{abstract}
Within a microscopic approach which takes into account RPA configurations,
the single-particle continuum and more complex $1p1h\otimes phonon$
configurations isoscalar and isovector M1 excitations for the unstable
nuclei $^{56,78}$Ni and $^{100,132}$Sn are calculated. For comparison,
the experimentally known M1 excitations in $^{40}$Ca and $^{208}$Pb
have also been calculated. In the latter nuclei good agreement in the
centroid energy, the total transition strength and the resonance width
is obtained. With the same parameters we predict the magnetic excitations
for the unstable nuclei. The strength is sufficiently concentrated to
be measurable in radioactive beam experiments. New features are found
for the very neutron rich nucleus $^{78}$Ni and the neutron deficient
nucleus $^{100}$Sn.
\end{abstract}

\vspace{.75in}
\begin{flushright}
PACS Indices: 21.10.-k\\
21.60.-n\\
24.30.Cz
\end{flushright}
\baselineskip 20pt
\newpage
\section{Introduction}
There is a rich history in the study of M1 excitations in nuclei
(see ref.~\cite{RaFH} for a recent review). For more than two decades
the experimental effort has concentrated on stable
magic nuclei . Unfortunately,
in the valley of stability, the number of cases is quite limited which
rules out systematics of the quality we are used to from electric
resonances \cite{SpWa}.  With the availability of radioactive beam
facilities this may change since far from stability
new nuclei become available with
similar shell characteristics as the stable magic ones \cite{Kien}.
Aside from
systematics such nuclei are also of great astrophysical interest
(see references in \cite{Cruz}).

The aim of the present paper is to extend the microscopic models which
describe well the known resonances to the new region
accessible by radioactive beams.  As is well known, the RPA with
phenomenological Landau-Migdal interactions is able to reproduce the
excitation energies of known isoscalar and isovector M1 excitations
with a universal set of spin-interaction parameters $g$ and $g'$.
Including effective spin g-factors $\sim 0.7-0.8$ the observed total
transition strengths are also reproduced \cite{RiSp,TkBK,BoTk,BoTF}.
The RPA fails,
however, to account for the resonance width which, especially in heavy
nuclei, is caused by a coupling to more complex configurations.
For the M1 case there are several approaches for including such
configurations
\cite{KPYV,CSWS,KaTk}. Concerning the quenching of M1 strength
the most reliable approach seems to be the use of phenomenological
values for the g-factors. By now, these are very well establish from
a variety of experiments \cite{BoTF,PjFa} and are universal for all
nuclei measured. For predictions in unstable nuclei knowledge of
global interaction parameters and $g$-factors is of crucial
importance.

Most realistically, the more complex configurations are described
as one-particle one-hole states coupled to low-lying surface
vibrations ($1p1h\otimes phonon$ states) using known parameters.
Recently a formalism for including such
configurations in conjunction with the single-particle continuum
has been
successfully used in the description of giant electric resonances
\cite{KTU1,KST1}. In particular, the widths are reproduced
satisfactorily. We shall employ this approach here in the
study of M1 transitions in unstable nuclei.

The theory will be briefly described in sect.~2. In sections 3. and 4.
calculations for the known M1 states in $^{48}$Ca and $^{208}$Pb are
presented which show the quality of the theoretical results. The same
interaction parameters, effective charges and a standard mean field of
the Woods-Saxon type are then used to predict spectra for unstable nuclei.

\section{Theory}

The approach which we shall use has been developed in
the framework  of the consistent Green's function method
and is based on Migdal's Theory of Finite Fermi Systems (TFFS)
\cite{Migd}.
The main physical idea is to include $1p1h\otimes phonon$ configurations
instead of "pure" 2p2h ones and thus to make use of the fact that,
in magic nuclei, the squared phonon creation amplitude is a
small parameter
$g^2<1$. This enables one to restrict the calculation to particle-phonon
coupling terms of order $g^2$.
Furthermore only a small number of phonons of maximum amplitude
(maximal $g^2$) need to be considered, \ie the most collective low-lying
ones. This makes it possible to use known parameters of the TFFS which
determine the local effective interaction and quasiparticle charges
(taking phonons into account explicitly gives
non-local contributions). In addition, a small number of phonons
greatly reduces the computational effort. This physical assumption
has been confirmed in
several calculations of $M1$ excitations \cite{KaTk},
the giant dipole resonance \cite{KTU1,KST1} as well as
$E2,E0$ resonances \cite{KST2}.

The theory is formulated, most efficiently, in coordinate space. For
an applied external field $V^0$ with frequency $\omega$,
the change in density $\delta\rho= \rho-\rho_0$ from equilibrium
is given by
\beq
\delta\rho(\br,\omega)=-\int d\br'A(\br,\br',\omega)
\bigl (e_qV^0(\br')-{\cal F}(\br')\delta\rho(\br',\omega)\bigr )
\label{eq:drho}
\eeq
and the frequency distribution determines the excitation spectrum of
the system. Here $e_q$ denotes the local quasiparticle charge and
${\cal F}$ the quasiparticle interaction. As discussed above, these can
be taken from the TFFS. The generalized propagator $A$ contains
the RPA part as well as the $1p1h\otimes phonon$ configurations and was
derived in ref.~\cite{Tsel}  in a representation of the discrete
single-particle basis $\{\tilde\phi_\lambda,\tilde\epsilon_\lambda\}$.
The tilde indicates that these states
have been corrected for mean field contributions from the particle-phonon
coupling,  already included in phenomenological potentials of the
Woods-Saxon type. Since these contributions are included explicitly they
have to be removed from the single-particle potential to avoid double
counting. To include single-particle emission, the discrete propagator
$A_{1234}$ has to be augmented by a continuum part and takes the form
\beq
A(\br,\br',\omega)=\tilde A^{RPA}_{cont}(\br,\br',\omega)
+\sum_{1234}\bigl [A_{1234}(\omega)-\tilde A^{RPA}_{1234}
(\omega)\delta_{13}\delta_{24}
\bigr ]\tilde \phi^*_1(\br)\tilde\phi_2(\br)
\tilde\phi_3(\br')\tilde\phi^*_4(\br')
\label{eq:Prop}
\eeq
where $\tilde A^{RPA}_{cont}$ is the continuum RPA propagator
(for details see ref.~\cite{ShBe,SaFK}).

The M1 transition probabilities between the ground- and excited states
are determined by the strength function
\beq
{dB(M1)\over d\omega}=\sum_n|\bra{n}e_qV^0\ket{0}|^2
\delta(\omega-\omega_n).
\eeq
Via the optical theorem this is related to $\delta\rho$ as
\beq
{dB(M1)\over d\omega}=-{1\over \pi}Im\int d\br e_qV^0(\br)
\delta\rho(\br,\omega)
\eeq
and  the solutions of Eq.~(\ref{eq:drho}) determine $dB(M1)/d\omega$.

The summation over single-particle states in Eq.~(\ref{eq:Prop}) is
usually performed for two shell above and below the Fermi level.  Aside
from providing a finite width above the continuum threshold our method
has the distinct advantage, over discretized calculations of this type,
that the matrix dimension still remains manageable.

A major difference between the approach presented above and similar
approaches \cite{Solo,BoBr,WaML} is the consistent inclusion of
ground state
correlations (GSC) beyond the RPA. There are two kinds of such $2p2h$
(to be more exact $1p1h\otimes phonon$ in our case) correlations. One is
GSCs without "backward going" diagrams containing the quasiparticle-phonon
interaction. The second one which includes these diagrams has new poles
which generate new states.
In the RPA, GSCs are caused by non-pole diagrams, only, which cannot
produce additional
excited states. In this sense our theoretical extension is qualitatively
different from the RPA. For M1 excitations in magic nuclei the consequences
have been discussed in Refs.~\cite{KaTk,KaTs} without taking into
account the
single-particle continuum. For the spin-saturated nuclei $^{16}$O and
$^{40}$Ca, where strong M1 excitations are not allowed because of
missing spin-orbit partners, one obtains M1 excitations which
describe the experimental data reasonably well.

\section{Approximations and Parameters}

The main physical approximations have been described above and in
\cite{KTU1,KST1}. Here we present  the parameters used for the present
calculations.

According to the selection rules only the spin-dependent part of the
Landau-Migdal interaction
\beq
{\cal F}=C_0(g+g'\btau\cdot\btau')\bsigma\cdot\bsigma'
\delta(\br_1-\br_2)
\label{eq:para1}
\eeq
enters in the M1 excitations. In order to predict the M1 states in the
unstable nuclei reliably we have slightly adjusted the $g'$ parameter to
obtain the well-known $1^+$ levels in  $^{48}$Ca. This gives $g'=0.86$
instead of $g'=0.96$ used previously \cite{BoTF,KaTk,KaTs,Tsel}
($C_0=300$ MeV fm$^{3}$).
For the parameter $g$ we have used $g=-0.05$.

The local quasiparticle charge was determined from \cite{Migd}
\bea
e_q^pV^{0p}&=&(1-\xi_l)\bj^p+[(1-\xi_s)\gamma^p+
\xi_s\gamma^n+1/2\xi_l-1/2]\bsigma^p\nonumber\\
e_q^nV^{0n}&=&\xi_l\bj^n+[(1-\xi_s)\gamma^n+
\xi_s\gamma^p-1/2\xi_l ] \bsigma^n
\eea
where $\gamma^p=2.79\mu_0$, $\gamma^n=-1.91\mu_0$, $\mu_0=e\hbar/2m_pc$ and
\beq
\xi_s^p=\xi_s^n=0.1,\qquad \xi_l^p=\xi_l^n=-0.03
\label{eq:para2}
\eeq
as obtained earlier \cite{Baue,BoTF,KaTk,KaTs,Tsel}. These values yield
for the spin local
charge $e^p_q=0.64\gamma^p$ and $e_q^n=0.74\gamma^n$.

The low-lying collective phonons (listed in Table 1 for the various nuclei
considered) which are  taken into account in the $1p1h\otimes phonon$
coupling have been calculated within RPA using the following known
Landau-Migdal parameters
\bea
f_{in}=-0.002,\quad f'_{ex}=2.30,\quad f'_{in}=0.76\nonumber\\
g=-0.05,\quad g'=0.96,\quad C_0=300 MeV fm^3
\eea
for all nuclei. Since the spin-dependent contributions to the low-lying
phonon spectrum is small, the difference between $g'=0.86$ and $g'=0.96$
has no noticeable consequences for the results.  The parameters $f_{ex}$
has been fitted to available phonon energies  and ranges from -5.0
to -3.5 which is not far from the value $f_{ex}=-3.74$ used earlier
\cite{KTU1,KST1,KST2}.

The single-particle states have been calculated from a standard Woods-Saxon
potential \cite{Chep}. To get good agreement with measured single-particle
energies which are available for $^{48}$Ca, $^{208}$Pb, $^{56}$Ni
\cite{Lean}, $^{132}$Sn \cite{Bjor,Lean} and, to some extent,
for $^{100}$Sn \cite{Lean,Doba} the
well depth of the central part of the potential has been adjusted,
changing the depth parameter by less than 5\%. A list of
single-particle energies, thus obtained, is given in Table 2.
Another check is the energy of the
$g_{9/2}(p)\to g_{7/2}(n)$ Gamow-Teller transition in $^{100}$Sn. We
find a transition energy of 7.6 MeV which coincides with the results
of Ref.~\cite{Doba}.

As mentioned above, the coupling of single-particle states to phonons
gives a contribution to the observed single-particle energies.
When adjusting the Woods-Saxon potential to empirical energies such
contributions are already accounted for. In order to avoid double
counting the particle-phonon contributions should be removed. This leads
to a "refined" basis $\{\tilde\phi_\lambda,\tilde\epsilon_\lambda \}$.
The corrected energies $\tilde\epsilon_\lambda$ are obtained by solving
the non-linear equations
\bea
\tilde\epsilon_\lambda&=&\epsilon_\lambda-
{\cal M}_{\lambda\lambda}(\epsilon_\lambda)\nonumber\\
 .&=&\epsilon_\lambda-{1\over 2j+1}\sum_{s,\lambda'}|
\bra{\lambda}g_s\ket{\lambda'}|^2\biggl\{
{1-n_{\lambda'}\over \epsilon_\lambda-\tilde\epsilon_{\lambda'}-\omega_s}
+{n_{\lambda'}\over \epsilon_\lambda-\tilde\epsilon_{\lambda'}+\omega_s}
\biggr \}
\eea
where $\lambda\equiv\{n,l,j\}$.

Finally, in order to account for instrumental resolution in the
experiment we have included an energy averaging parameter $\Delta$.

\section{Results and Discussion}
The results of our calculations are presented in Table 3 and in
Figs.~1-10. The continuum RPA results are denoted by {\it 1p1h+continuum}
while {\it 1p1h+2p2h+continuum} indicates the results including the
$1p1h\otimes phonon$ contributions. To elucidate the role of the additional
ground state correlations we quote results with and without ($gs(\pm)$)
in the figures. The mean energies listed in Table 3 are defined as
\beq
\bar E={\sum_iE_iB_i(M1\uparrow)\over\sum_iB_i(M1\uparrow)}.
\eeq
Fig.~1 shows the calculated strength distribution for $^{48}$Ca. In the
continuum RPA (dotted line) this is dominated by the $1f_{7/2}(n)\to
1f_{5/5}(n)$ transition.
After inclusion of more complicated configurations the
measured centroid energy of 10.23 MeV is quite well reproduced
($\bar E^{th}=10.36$ MeV) while the calculated transition strength is
somewhat larger than in the experiment. The additional ground
state correlations have no noticeable effect. In $^{208}$Pb the two
particle-hole transitions $1h_{11/2}(p)\to 1h_{9/2}(p)$ and $1i_{13/2}(n)
\to 1i_{11/2}(n)$ form an isoscalar and isovector transition within the
RPA. Taking into account the more complex configurations
we obtain reasonable agreement with experiment for both the isoscalar
$1^+$ level at $E_{exp}$=5.85 MeV and for the isovector M1 resonance
(Fig.~2). A rather
strong quenching has been obtained so that our value $\sum B(M1\uparrow)=
11.6 \mu_0^2$ is less than the experimental value of $15.6 \mu_0^2$
\cite{Las4}. The calculated centroid
energy ($\bar E^{th}=7.74$ MeV) is slightly above the measured one
($\bar E^{ex}=7.3$ MeV). The width which can be
deduced from Fig.~2 is in good agreement with the experimental value of
$\sim$ 1 MeV. The effect of additional GSC is more pronounced than in
$^{48}$Ca: it increases the strength in the energy interval 9.9-15.5 MeV
by about $3\mu_0^2$ (see Table 3).

Table 3 and Fig.~3 give the results for $^{56}$Ni \cite{Cruz}.
As with $^{208}$Pb
this is a non-spin saturated nucleus with the $1f_{7/2}$ shell filled
for protons and neutrons. Thus an isoscalar and isovector state can be
formed. The influence of the $1p1h\otimes phonon$ states is quite noticeable
reducing the strength in the peak by more than a factor of two. As can be
seen from Fig.~3 and, in particular, from  Fig.~8 (here the averaging
parameter is 20 keV only) there is no fragmentation width for the resonance
peak. The latter result also seems to hold for $^{132}$Sn (Figs.~6 and 10).
For $^{56}$Ni we confirm the result of ref.~\cite{MDSW} where no
fragmentation width was found in a continuum RPA calculation.

We now come to the "very exotic" nuclei $^{78}$Ni and $^{100}$Sn.
In the former the $1f_{7/2}(p)$ and
$1g_{9/2}(n)$ are occupied while in the latter both $1g_{9/2}$ shells are
full. As seen in Fig.~4 $^{78}$Ni has an asymmetric resonance shape with
a width of $\sim$ 1 MeV. The M1 resonance in the very neutron deficient
$^{100}$Sn is split into two major peaks with $\bar E_1=9.8$ MeV and
$\bar E_2=10.5$ MeV (Fig.~5). The low-lying isoscalar level is spit also.
These features are caused by the inclusion of the $1p1h\otimes phonon$
configurations and for $^{100}$Sn also by the very small value of the
proton binding energy $B_p=2.91$ MeV.

\section{Conclusion}

Within a microscopic approach which includes the continuum RPA as well as
$1p1h\otimes phonon$ configurations we have calculated isoscalar and
isovector M1 excitations in the unstable nuclei $^{56,78}$Ni and
$^{100,138}$Sn. To judge the quality of our predictions we have also
given results for the "known" M1 excitations in $^{48}$Ca and $^{208}$Pb.
For the latter good agreement with experiment in the centroid
energy, the transition strength and the width is obtained. This, together
with the fact that we have used known universal parameters for the
effective interaction and the local M1 charge, gives
us confidence that the predictions for the unstable nuclei are realistic.

The location of the isovector M1 resonances agrees rather well with the
RPA calculations.
In all cases studied, we find a considerable influence of the
particle-phonon coupling on the width and the
strength distributions. As compared with the RPA there is noticeable
strength in the high-energy  tails and a decrease of strength in the
resonance region.
The damping effects are particularly pronounced in the extremely neutron
rich nucleus $^{78}$Ni as well as the neutron deficient $^{100}$Sn. We
did not find a simple A-dependence of the resonance widths.
All these features depend delicately on the interplay
between the shell structure and the low-lying vibrational modes.

For approaches, like the one used here, which are based on a phenomenological
single-particle scheme and the effective Landau-Migdal interaction it is
important to experimentally confirm the results presented here. In particular,
it would give reassurance of the universality (A-independence) of the
Migdal parameters (\ref{eq:para1}) and (\ref{eq:para2}) and hence would
allow to generalize the present approach to non-magic nuclei and to use it
for many nuclei far from stability.

\vspace{2.0cm}

\bce
{\bf Acknowledgement}
\ece

\noindent

This work was supported by the German-Russian exchange program in part
by NSF grant PHY-89-21025. Two of us, S.K. and G.T. express their gratitude
for warm hospitality of the IKP at the Forschungszentrum J\"ulich. Useful
discussions with Dr. S. Dro\.zd\.z and Profs. S. Krewald, F. Osterfeld,
A. Richter, P. Ring, K. Sistemich and D. Zawischa are gratefully
acknowledged.

\newpage
\vspace{.25in}
\parindent=.0cm            

\newpage
\begin{center}
{\bf\large Table 1} : {\it \large
Characteristics of the low-lying phonons used in
the calculations}

\vspace{1cm}
\begin{tabular}{|l|l|l|l|l|l|l|l|l|} \hline
 J$^{\pi}$ &E, &B(EL)${\uparrow}$, &J$^{\pi}$ &E, &B(EL)${\uparrow}$,
&J$^{\pi}$ &E, &B(EL)${\uparrow}$, \\
   &MeV &e$^{2}$fm$^{2L}$ & &Mev &e$^{2}$fm$^{2L}$ & &MeV &e$^{2}$
fm$^{2L}$ \\ \hline
\multicolumn{3}{|c}
{\bf$^{\bf{100}}$Sn} &
\multicolumn{3}{|c|}
{\bf$^{\bf{132}}$Sn}  &
\multicolumn{3}{c|}
{\bf $^{\bf{48}}$Ca}    \\ \hline
 2$^ {+}$ &2.86 &331  &2$^{+}$ &4.06 &942 &2$^{+}$ &3.83 &81.6
 \\ \hline
 3$^{-}$ &3.56 &5.33 10$^{3}$ &3$^{-}$ &4.34 &5.0 10$^{4}$ &3$^{-}$ &4.50
 &1.12 10$^{4}$ \\ \hline
 3$^{-}$ &5.03 &5.99 10$^{3}$ &5$^{-}$ &4.91 &3.46 10$^{6}$&
\multicolumn{3}{c|}
{\bf $^{\bf{208}}$Pb}  \\ \hline
 5$^{-}$ &3.57 &2.29 10$^{7}$ &5$^{-}$ &5.61 &5.07 10$^{6}$
 &3$^{-}$ &2.61 &3.44 10$^{5}$ \\ \hline
 4$^{+}$ &3.70 &1.02 10$^{5}$ &5$^{-}$ &6.06 &5.43 10$^{6}$
&2$^{+}$ &4.07 &2.33 10$^{3}$  \\ \hline
 4$^{+}$ &4.09 &1.68 10$^{5}$ &5$^{-}$ &6.85 &2.10 10$^{7}$
&2$^{+}$ &9.71 &3.30 10$^{3}$ \\ \hline
 6$^{+}$ &3.90 &9.16 10$^{7}$ &4$^{+}$ &4.24 &1.80 10$^{6}$
&4$^{+}$ &4.34 &6.64 10$^{6}$ \\ \hline
 6$^{+}$ &4.81 &2.65 10$^{9}$ &6$^{+}$ &4.75 &1.14 10$^{9}$
&6$^{+}$ &4.40 &1.64 10$^{10}$ \\ \hline
 6$^{+}$ &6.47 &6.36 10$^{8}$ &6$^{+}$ &5.42 &1.06 10$^{9}$
&5$^{-}$ &3.20 &2.60 10$^{8}$  \\ \hline
 6$^{+}$ &6.79 &1.50 10$^{9}$  &
\multicolumn{3}{c|}
{\bf$^{\bf{78}}$Ni}
&5$^{-}$ &3.70 &9.87 10$^{7}$  \\ \hline
\multicolumn{3}{|c|}
{\bf$^{\bf{56}}$Ni}
&2$^{+}$ &3.53 &145 & & & \\ \hline
 2$^{+}$ &2.73 &361      &4$^{+}$ &3.70 &8.28 10$^{4}$ & & & \\ \hline
3$^{-}$ &4.62 &1.28 10$^{4}$ &4$^{+}$ &4.67 &6.52 10$^{4}$
& & & \\ \hline
5$^{-}$ &6.35 &2.90 10$^{6}$ &4$^{+}$ &4.99 &4.26 10$^{4}$
& & & \\ \hline
4$^{+}$ &3.73 &2.27 10$^{5}$ &6$^{+}$ &4.00 &2.32 10$^{6}$
& & & \\ \hline
6$^{+}$ &5.18 &1.36 10$^{8}$ &6$^{+}$ &4.94 &4.17 10$^{7}$
& & & \\ \hline
 & &                         &3$^{-}$ &4.38 &8.32 10$^{3}$
& & & \\ \hline
 & &                         &5$^{-}$ &5.04 &1.18 10$^{6}$
& & & \\ \hline
\end{tabular}
\end{center}
\newpage
\begin{center}
{\bf\large Table 2} : {\it \large The single-particle levels for the
unstable nuclei used in the calculations}
\end{center}

\begin{table}[h]
\advance\leftskip by 2 cm
\advance\rightskip by 2 cm
\bigskip
{\offinterlineskip \tabskip=0pt \halign{\strut
\vrule \hfil \ #\ & \vrule \hfil \ #\ &
\vrule \hfil \ #\ & \vrule \hfil \ #\ &
\vrule \hfil \ #\ & \vrule#\cr
\noalign{\hrule}
& &\multispan 1 {\bf $^{132}$Sn}\hfil& &\multispan 1 {\bf $^{100}$Sn} &\cr
\noalign{\hrule}
            { {nlj} }&
            { {n} }&
            { {p} }&
            { {n} }&
            { {p} }&\cr
\noalign{\hrule}
     1s1/2& -37.36& -39.49& -45.81& -29.52&\cr
     1p3/2& -33.02& -34.78& -39.81& -24.33&\cr
     1p1/2& -31.61& -33.32& -38.07& -22.34&\cr
     1d5/2& -28.23& -28.32& -33.94& -19.12&\cr
     1d3/2& -24.47& -26.05& -30.25& -15.03&\cr
     2s1/2& -23.93& -25.36& -29.38& -14.06&\cr
     1f7/2& -21.27& -22.91& -25.64& -11.47&\cr
     2p3/2& -17.17& -17.67& -20.75&  -6.13&\cr
     1f5/2& -16.39& -17.65& -21.28&  -6.56&\cr
     2p1/2& -15.48& -15.91& -18.75&  -4.12&\cr
     1g9/2& -14.93& -15.45& -17.59&  -2.91&\cr
     1g7/2&  -9.64&  -9.70& -10.70&   3.45&\cr
     2d5/2&  -8.97&  -8.83& -12.73&   1.07&\cr
     3s1/2&  -7.55&  -7.10&  -9.82&   1.15&\cr
    1h11/2&  -7.53&  -6.96& -10.37&   0.65&\cr
    2d3/2&   -7.15&  -6.70&  -9.55&   4.15&\cr
    2f7/2&   -2.50&  -1.57&  -3.57&   8.75&\cr
    1h9/2&   -1.49&   0.25&  -1.25&  12.05&\cr
    3p3/2&   -1.36&   0.65&  -2.11&  10.45&\cr
    3p1/2&   -0.59&   1.15&  -1.11&  11.95&\cr
    1i13/2&  -0.25&  -0.14&  -2.40&  10.15&\cr
    3d5/2&    2.35&   &         &         &\cr
    2g9/2&    3.65&   &         &         &\cr
    2g7/2&    6.35&   &         &         &\cr
    2f5/2&    8.05&   0.45&   -0.96&  11.25&\cr
    1i11/2&   9.65&   &           &        &\cr
    3d3/2&   15.35&   &         &          &\cr
\noalign{\hrule}}}
\end{table}
\newpage
\begin{center}
{\bf\large Table 2} : {\it \large The single-particle levels for the
unstable nuclei used in the calculations}
\end{center}

\begin{table}[h]
\advance\leftskip by 2 cm
\advance\rightskip by 2 cm
\bigskip
{\offinterlineskip \tabskip=0pt \halign{\strut
\vrule \hfil \ #\ & \vrule \hfil \ #\ &
\vrule \hfil \ #\ & \vrule \hfil \ #\ &
\vrule \hfil \ #\ & \vrule#\cr
\noalign{\hrule}
& &\multispan 1 {\bf $^{56}$Ni}\hfil & &\multispan 1 {\bf $^{78}$Ni} &\cr
\noalign{\hrule}
            { {nlj} }&
            { {n} }&
            { {p} }&
            { {n} }&
            { {p} }&\cr
\noalign{\hrule}
       1s1/2& -44.34& -37.19& -35.05& -41.39&\cr
       1p3/2& -33.68& -26.12& -28.73& -34.73&\cr
       1p1/2& -32.99& -24.39& -26.44& -32.42&\cr
       1d5/2& -27.29& -20.74& -21.74& -27.19&\cr
       2s1/2& -22.79& -15.98& -17.92& -22.77&\cr
       1d3/2& -21.33& -14.73& -18.75& -22.45&\cr
       1f7/2& -16.83& -7.92&  -14.13& -18.88&\cr
       2p3/2& -10.69& -3.58&  -9.61&  -13.32&\cr
       1f5/2& -10.04& -1.08&  -7.03&  -11.41&\cr
       2p1/2& -9.80&  -2.06&  -7.54&  -10.96&\cr
       1g9/2& -7.51&  -0.023& -5.98&  -9.87&\cr
       2d5/2& -3.20&   2.95&  -1.95&  -4.02&\cr
       3s1/2& -2.12&   3.85&  -1.14&  -1.87&\cr
       2d3/2& -0.11&   5.85&  -0.39&  -0.50&\cr
       1g7/2&   &        &    -0.76&    &\cr
      1h11/2&   &        &    -0.31&    &\cr
       3p3/2&   &        &     1.35&    &\cr
       2f7/2&   &        &     4.75&    &\cr
       3p1/2&   &        &     5.35&    &\cr
       1h9/2&   &        &     7.65&    &\cr
      1i13/2&   &        &     7.85&    &\cr
       2f5/2&   &        &     8.15&    &\cr
\noalign{\hrule}}}
\end{table}
\newpage
\footnotesize
\pagestyle{empty}
\noindent
\vspace{0.2cm}
{\bf\large Table 3} : {\it \large Characteristics of M1 excitations}  \\
\vspace{1.0cm}
{\normalsize ($E_{is}$,$E_{iv}$,$\bar{E}$
are given in MeV,$B(M1)\uparrow$ and
$\sum  B(M1)\uparrow$ in $\mu_0^2$)} \\
\vspace{1.5cm}
\scriptsize
\begin{tabular}{*{9}{|c}|}\hline
& \mc{4}{c|}{}
& \mc{4}{c|}{$1p1h+2p2h+cont.$}\\
& \mc{4}{c|}{\raisebox{1.5ex}[1.5ex]{$1p1h+cont.$}}
& \mc{4}{c|}{without $2p2h\,GSC$} \\ \hline
& & & & $\sum  B(M1)\uparrow$
& & & & $\sum  B(M1)\uparrow$    \\
&\raisebox{1.5ex}[1.5ex]{$E_{is}$}
&\raisebox{1.5ex}[1.5ex]{$B(M1)\uparrow$}
&\raisebox{1.5ex}[1.5ex]{$E_{iv}$}
&(interval)
&\raisebox{1.5ex}[1.5ex]{$E_{is}$}
&\raisebox{1.5ex}[1.5ex]{$B(M1)\uparrow$}
&\raisebox{1.5ex}[1.5ex]{$\bar{E}$}
&(interval)  \\ \hline
& & & & & & & $E_{iv}=$ & \\
$^{48}Ca$ & & &10.68 &8.64 & &
&10.35 &6.55 \\
& & & &($\sim10.68$)& & & &($\sim10.35$)\\
Figs.~1,7& & & & & & & &8.24 \\
& & & & & & & &(9-19) \\ \hline
&5.63 &0.49 &7.97 &18.16 &5.72 &0.84
&7.66 &11.87 \\
$^{208}Pb$ & & & &($\sim7.97$)& & & &(6.3-8.7) \\
Fig.~2& & & &(5-15.5) & & & &(8.7-9.9) \\
& & & & & & & &4.98 \\
& & & & & & & &(9.9-15.5) \\
& & & & & & & &19.09 \\
& & & & & & & &(5.0-15.5) \\ \hline
 &6.56 &0.31 &10.12 &11.31 &6.7 &0.24
&9.93 &4.8 \\
$^{56}Ni$ & & & &($\sim10.12$)& & & &(8.5-11.0) \\
& & & &11.43 & & & &5.58 \\
Figs.~3,8& & & &(5.5-15) & & & &(11-15) \\
& & & &11.73 & & & &10.62 \\
& & & &(6-45) & & & &(5.5-15) \\ \hline
&5.88 &0.66 &10.16 &16.0 &6.12 &0.64
&9.3 &10.2 \\
$^{78}Ni$ & & & &($\sim10.16$)& & & &(8.5-10.2) \\
& & & &16.7 & & & &6.8 \\
Figs.~4,9& & & &(6.5-15) & & & &(10.2-15.0) \\
& & & & & & & &17.64 \\
& & & & & & & &(5.5-15) \\ \hline
 &6.8 &1.5 &10.15 &14.1 &6.5 &0.3
&8.9 &1.3 \\
$^{100}Sn$ & & & &($\sim10.15$)& & & &(7-9.4) \\
& & & &15.6 &6.78 &1.4 &9.77 &6.6  \\
Fig.~5& & & &(6-13) &($\bar{E}=$ & & &(9.4-10.15) \\
& & & & &6.67) & & & \\
& & & & & & &10.5 &3.2 \\
& & & & & & &($\bar{E}=$ &(10.15-10.7)\\
& & & & & & &10.0) &3 \\
& & & & & & & &(10.7-13)     \\
& & & & & & & &15.8  \\
& & & & & & & &(6-13) \\ \hline
 &5.8 &0.4 &8.78 &19.0 &5.8 &0.4
&8.45 &13.2 \\
$^{132}Sn$ & & & &($\sim8.78$)& & & &($\sim8.45$)   \\
& & & &19.6 & & & &6.0 \\
Figs.~6,10& & & &(5.5-15) & & & &(9-15)\\
& & & & & & & &19.2      \\
& & & & & & & &(5.5-15)   \\ \hline
\end{tabular}

\newpage
\footnotesize
\pagestyle{empty}
\noindent
\vspace{1.0cm}
{\bf \large Table 3} : {\it \large (continuation)}\\
\scriptsize
\begin{tabular}{*{9}{|c}|}\hline
& \mc{4}{|c|}{$1p1h+2p2h+cont.$}
& \mc{4}{c|}{}  \\
& \mc{4}{|c|}{with $2p2h\,GSC$}
& \mc{4}{c|}{\raisebox{1.5ex}[1.5ex]{Experiment}} \\  \hline
& & & & $\sum B(M1)\uparrow$
& & & & $ B(M1)\uparrow$ or $\sum B(M1)\uparrow$ \\
&\raisebox{1.5ex}[1.5ex]{$E_{is}$}
&\raisebox{1.5ex}[1.5ex]{$B(M1)\uparrow$}
&\raisebox{1.5ex}[1.5ex]{$\bar{E}$}
&(interval)
&\raisebox{1.5ex}[1.5ex]{$E_{is}$}
&\raisebox{1.5ex}[1.5ex]{$B(M1)\uparrow$}
&\raisebox{1.5ex}[1.5ex]{$E_{iv}$ or $\bar{E}$}
&\mc{1}{r|}{(interval)} \\ \hline
& & & $E_{iv}=$ & &
& & $E_{iv}=$ & \\
$^{48}Ca$& & &10.36 &6.12 & & &10.32
&$3.9\pm0.3[31]$\\
& & & &($\sim10.36$)
& & & & \\
Figs.~1,7& & & &9.68 & & & &$5.3\pm0.6[31]$ \\
&& & &(9-19) & & & &(7.7-12.7) \\ \hline
&5.72 &0.84 &7.74 &11.57 &5.85 & $1.6\pm0.5[33]$ & 7.3
& $\approx 15.6 [29]$ \\
$^{208}Pb$& & & &(6.3-8.7) & & & & (6.7-8.4)\\
& & & &2.2 &5.85 & $1.01_{-0.13}^{+0.42}[32]$
& &  \\
Fig.~2& & & &(8.7-9.9)
& & & & \\
& & & &8.02 &5.85 &
& &  \\
&& & &(9.9-15.5)
& & $1.9^{+0.7}_{-0.4}[29]$ & & \\
&& & &22.6 &6.24 &
& &  \\
&& & &(5.0-15.5)
& & & & \\ \hline
&& & & &
\mc{4}{c}{}   \\
$^{56}Ni$&6.7 &0.24 &10.02 &5.5 &\mc{4}{c}{} \\
& & & &(8.5-11.0)
&\mc{4}{c}{}   \\
Figs.~3,8& & & &5.94
&\mc{4}{c}{}   \\
& & & &(11-15)
&\mc{4}{c}{}   \\
& & & &11.68
&\mc{4}{c}{}   \\
& & & &(6-15)
&\mc{4}{c}{}   \\ \cline{1-5}
& & & & &
\mc{4}{c}{}   \\
$^{78}Ni$&6.12 &0.64 &9.3 &9.5
&\mc{4}{c}{}   \\
& & & &(8.5-10.2)
&\mc{4}{c}{}   \\
Figs.~4,9& & & &7.6
&\mc{4}{c}{}   \\
& & & &(10.2-15.0)
&\mc{4}{c}{}   \\
& & & &17.74
&\mc{4}{c}{}   \\
&& & &(5.5-15)
&\mc{4}{c}{}   \\
\cline{1-5}
\end{tabular}

\newpage
\normalsize
\begin{center}
{\bf Figure Captions}
\end{center}

      Fig. 1.   The  M1  strength function (in MeV) for $^{48}$Ca with

{}~~~~~~~~~~~~~ an averaging parameter   $\Delta = 100$ keV.

      Fig. 2.   Same as in Fig. 1 but for  $^{208}$Pb.

      Fig. 3.   Same as in Fig. 1 but for $^{56}$Ni.

      Fig. 4.   Same as in Fig. 1 but for $^{78}$Ni.

      Fig. 5.   Same as in Fig.~1 but for $^{100}$Sn.

      Fig. 6.   Same as in Fig.~1 but for $^{132}$Sn.

      Fig. 7.   The M1 strength function for $^{48}$Ca, $\Delta= 10$ keV.

      Fig. 8.   The M1 strength function for $^{56}$Ni, $\Delta= 20$ keV.

      Fig. 9.   The M1 strength function for $^{78}$Ni, $\Delta= 20$ keV.

      Fig. 10.  The M1 strength function for $^{132}$Sn, $\Delta= 250$ keV.

\end{document}